\documentclass[aps,prl,showpacs,twocolumn]{revtex4}
\usepackage{psfig}
\usepackage{amssymb}

\begin{document}
\title{Depolarization induced by subwavelength metal hole arrays}

\author{C. Genet, E. Altewischer, M.P. van Exter and J.P. Woerdman}

\affiliation{Huygens Laboratory, Leiden University, P.O. Box
9504,\\ 2300 RA Leiden, The Netherlands}

\begin{abstract} We present a symmetry-based theory of the
depolarization induced by subwavelength metal hole arrays. We
derive the Mueller matrices of hole arrays with various symmetries
(in particular square and hexagonal) when illuminated by a
finite-diameter (e.g. gaussian) beam. The depolarization is due to
a combination of two factors: (i) propagation of surface plasmons
along the surface of the array, (ii) a spread of wave vectors in
the incident beam.
\end{abstract}

\maketitle

\noindent The transmission of a metal film perforated with a
two-dimensional periodic array of subwavelength holes is
extraordinary large due to the resonant excitation of surface
plasmons (SPs) \cite{EbbesenNature1998}. Recently, the
polarization properties of this enhanced transmission have drawn
considerable attention
\cite{AltewischerNature2002,MorenoarXiv2003,
AltewischerJOSAB2003,PapakostasPRL2003,ElliottPrePrint2003,SarrazinPrePrint2003}.
A particularly intriguing polarization property is that
illuminating the array by a fully polarized beam at normal
incidence leads to important reduction of the degree of
polarization in transmission, corresponding to a strongly
space-variant character of the polarization state of the output
beam \cite{AltewischerJOSAB2003}. This seems in clear contrast
with what could be guessed from general symmetry arguments. For
plane wave illumination, a periodic structure can theoretically
only transform a pure input state of polarization (SOP) to another
pure output SOP (see below). Thus, the observed depolarization
must be critically related to additional spatial degrees of
freedom of the system. This is exactly what excitation of SPs can
potentially yield since SPs propagate in distinct directions
related to the symmetry of
the hole array.\\
\hspace*{5mm}The aim of this Letter is to provide a proper
theoretical framework describing these depolarization effects.
Formally, spatial symmetries of arrays within the context of SP
excitations must be confronted with photon polarization SU(2)
symmetry, a context where Mueller algebra is the natural tool
\cite{SchmiederJOSA1969,Kliger}. The central issue of our work is
that implementations of symmetry operations have to account for
the fact that SPs are a source of effective spatial dispersion at
the arrays interfaces but only so if they are excited by a
finite-diameter beam (e.g. a gaussian beam). As we will discuss,
it is the strong angular dependence related to such a spatially
dispersive transmission that can lead to
sizeable depolarization.\\
\hspace*{5mm}Any two-dimensional pattern can only match five
possible lattices: the five 2D Bravais lattices \cite{Lovett}. In
order to complete such patterns into 2D crystal structures, a
primitive cell is associated with each lattice point. The required
symmetry compatibility of the primitive cell and the Bravais
lattice leads to the well-known seventeen 2D point groups. Here we
restrict ourselves to circular holes on a Bravais lattice, in
other words to the simplest primitive cell with full symmetry.
Therefore, the point group symmetry of our arrays is reduced to
the spatial symmetry of the chosen Bravais lattice which is a
$C_{nv}$ group, where $n$ denotes rotations by $2\pi /n$ radians
about the origin (with $n=1,2,3,4$ or $6$) and $v$
refers to mirror symmetries (if allowed).\\
\hspace*{5mm}The optical transmission of a given array under
general conditions of illumination is fully described by the
transmission matrix which relates, at a specific wavelength
$\lambda$, the transmitted field ${\bf E}^{\rm out}$ to the
incoming field ${\bf E}^{\rm in}$. In the paraxial approximation
used throughout this Letter, far field angles
$\left(\theta_{x},\theta_{y}\right)$, collected in a 2-column
vector ${\bf \theta}$, define paraxial fields as two independent
complex numbers forming a spinor ${\bf E}=(E_{1}, E_{2})$ in the
chosen $(1,2)$-basis.
\begin{figure}[tbh]
\centerline{\psfig{figure=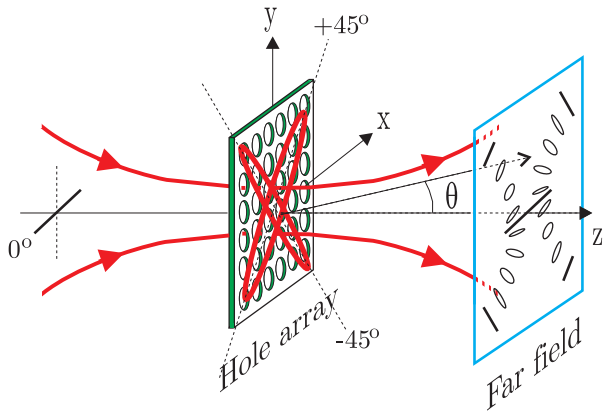,width=8cm}}
\caption{Cartoon-like representation of a hole array, chosen here
as a square array with circular holes, illuminated by a
finite-diameter (gaussian) beam. The input beam, chosen here as a
pure horizontal ($0^{\rm o}$) SOP, excites SPs along the principal
diagonals of the array (note that the reciprocal lattice coincides
with the direct lattice). On the backside of the array, the
corresponding near field pattern is shown with SPs propagating and
polarized along the $+45^{\rm o}/-135^{\rm o}$ and $+135^{\rm
o}/-45^{\rm o}$ lobes, onto which the input SOP can thus be
projected. The central region corresponding to the incident beam
remains polarization isotropic but when moving radially outward,
the input horizontal polarization is modified through elliptically
polarized states towards $\pm 45^{\rm o}$-linear polarizations due
to the longitudinal character of the SPs. As symbolized in the
figure, the output SOP becomes space-variant, the variation in
polarization states being parametrized by the far field angles
$\theta=(\theta_{x},\theta_{y})$. This 2D far field pattern is
Fourier related to the sketched near field pattern with
polarization directions remaining unmodified. Therefore, the
polarization directions observed in the far field, and more
particularly at large angles, are (maybe somewhat surprisingly)
associated with directions along which corresponding SPs {\it do
not} propagate, since a wide near field corresponds to a narrow
far field. The same general picture remains valid with an input
circular SOP. In this case, the input SOP passes through various
elliptically polarized states to reach the linear SOPs along the
diagonals. Extension of this description to an hexagonal array is
straightforward.}
\end{figure}
The transmission matrix ${\bf t}\left(\lambda ; \theta\right)$ of
the zero-order diffracted beam in the far field is therefore a
$2\times 2$ matrix
\begin{eqnarray}
{\bf E}^{\rm out}\left(\lambda ; \theta\right)={\bf
t}\left(\lambda ; \theta\right) {\bf E}^{\rm in}\left(\lambda ;
\theta\right).
\end{eqnarray}
From this input-output description, spatial symmetries of the
chosen array define orthogonal transformation matrices
$\mathcal{B}$ that leave the transmission matrix unchanged via
$\mathcal{B}^{\dagger}{\bf t}\left(\lambda ;
\theta\right)\mathcal{B}={\bf t}\left(\lambda ;
\mathcal{B}\theta\right)$. For the case of plane wave illumination
at arbitrary angle $\theta$, spatial symmetry operations therefore
merely relate transmission matrix elements with various angular
arguments. It is only when one resorts to angularly integrated
expressions, i.e. to illumination by a rotationally symmetric beam
(as for instance a gaussian beam at normal incidence), that one
recovers simple symmetry-based relations between
input and output SOPs of a given beam.\\
\hspace*{5mm}Depolarization effects are generally most efficiently
addressed when one resorts to the Stokes parameters $({\rm S}_{0},
{\rm S}_{1},{\rm S}_{2},{\rm S}_{3})$. These parameters, together
with the use of the associated Mueller algebra, define the natural
theoretical habitat of polarization properties in connection with
symmetry arguments \cite{SchmiederJOSA1969,Kliger}. The Stokes
parameters are real-valued and represent four intensity
measurements on a light beam: ${\rm S}_{0} = \left\langle
I\right\rangle $ corresponds to the total intensity in the beam,
as measured without any polarization selection, and ${\rm
S}_{1}=\left\langle I_{0^{\rm o}}-I_{90^{\rm o}} \right\rangle$,
${\rm S}_{2}=\left\langle I_{45^{\rm o}}-I_{-45^{\rm o}}
\right\rangle$, ${\rm S}_{3}=\left\langle
I_{\sigma^{+}}-I_{\sigma^{-}} \right\rangle$ represent three
balanced intensity measurements where $0^{\rm o}, 90^{\rm o}$ and
$\pm 45^{\rm o}$ refer to orientations of a linear analyzer and
$\sigma^{+} ( \sigma^{-})$ to a right(left)-handed circular
analyzer. The brackets $\left\langle \cdots \right\rangle$ stand
for averaging over the spread of the wave vectors of the beam,
i.e. an integration over the far field angles (see Fig. 1). A
similar integration in time has no effect, as the range of input
frequencies is assumed to be much smaller than the SP resonance
structure. The degree of polarization $\Pi$ of the beam is given
by $ \Pi = [({\rm S}_{1}^{2}+{\rm S}_{2}^{2}+{\rm S}_{3}^{2}) /
{\rm S}_{0}^{2}]^{1/2}$. For unpolarized light, ${\rm S}_{1},{\rm
S}_{2},{\rm S}_{3}$ components are ``averaged-out'' by the angular
integration so that $\Pi =0$. For a fully polarized beam, $\Pi =1$
whereas for partially polarized light, $0<\Pi <1$.\\
\hspace*{5mm}Collecting the Stokes parameters in a four-vector
${\bf S}=({\rm S}_{0},{\rm S}_{1},{\rm S}_{2},{\rm S}_{3})$,
transformation by the array from input to output Stokes vectors
${\bf S}^{{\rm out}}={\bf M}{\bf S}^{{\rm in}}$ is given by the
$4\times 4$ real-valued Mueller matrix ${\bf M}$ when assuming an
arbitrary input SOP with $\Pi = 1$. This matrix completely
describes all possible changes of this initially pure SOP due to
transmission through the array; it is directly related to the
transmission matrix ${\bf t}$ appearing in Eq. (1). The global
structure of the ${\bf M}$ matrix can be derived after we expand
input and output Stokes vectors in terms of SU(2) Pauli matrices
generators $(\sigma_{0}, \sigma_{1}, \sigma_{2}, \sigma_{3})$.
This expansion fixes the SU(2) decomposition of Mueller matrix
components to
\begin{eqnarray}
{\rm M}_{{\alpha\beta}}=\frac{1}{2}\left( {\cal
P}_{ijkl}(\sigma_{\alpha})_{ji} (\sigma_{\beta})_{kl}\right),
\end{eqnarray}
where we used summation over repeated indices. This decomposition
is based on the components ${\cal P}_{ijkl}=\left\langle
t_{ik}\left(\lambda ; \theta\right)t_{jl}^{\star}\left(\lambda ;
\theta\right)\right\rangle$ of the $4^{\rm th}$-rank tensor ${\cal
P}=\left\langle {\bf t} \otimes {{\bf t}}^{\star} \right\rangle$,
the star symbol denoting complex conjugation and the
``circled-cross'' symbol a tensorial product on ${\bf
t}$-matrices. Symmetry properties of the Mueller matrix will
result from conjunction between spatial symmetries of the lattice,
as implemented at the level of the averaged ${\cal P}$ components,
and the SU(2) symmetry of Pauli matrices. We assume a fully
polarized input beam, i.e. $\Pi ^{\rm in} = 1$ (see below for a
discussion of the general case), so that the average is performed
over the far field angles present in the input beam. More
precisely, this average corresponds to an intensity-weighted
integration over the input angles via $\left\langle \cdots
\right\rangle \equiv \int{\rm d}\theta\cdots
I^{\rm in}\left(\lambda ; \theta\right)$. \\
\hspace*{5mm}The simple case of plane wave illumination of the
array at normal incidence only retains the ${\bf \theta}={ 0}$
value in the averaging process. It is then straightforward to show
that for square ($C_{4v}$) and hexagonal ($C_{6v}$) spatial
symmetries, both the transmission matrix and, as immediately seen
from Eq. (2), the Mueller matrix are proportional to the identity
matrix. So we find, not surprisingly, perfect polarization
preservation for transmission through such arrays. For rectangular
($C_{2v}$) symmetry, the SOP may change (corresponding to
birefringence and/or dichroism) but the degree of polarization is
conserved. The situation becomes fundamentally different when the
array is illuminated with a finite-diameter beam and when, at the
same time, SPs are excited at the array interfaces. Eq. (2)
implies that these two aspects have to be {\it combined} in the
transmission process, as we explain now. \\
\hspace*{5mm}First, the finite-diameter of the incident beam
corresponds to a spread of incoming wave vectors which calls for
the averaging in the definition of the Stokes parameters and the
${\cal P}$ tensor. Second, SP excitations stick to preferred
propagation directions on the array interfaces according to the
well-known dispersion relation which restricts the propagation of
SP waves on the reciprocal lattice of the array
\cite{GhaemiPRB1998}. SP excitations are then possibly {\it
delocalized} beyond the excitation spot of the incident beam, the
condition being that the transverse coherence length of the input
beam at the surface of the array should be smaller than the SP
coherence length, which can among others be estimated from
spectral widths of transmission peaks associated to SP resonances
\cite{AltewischerJOSAB2003}. Identifiable propagation axes for SPs
induce an effective spatially dispersive response of the
illuminated array; this is analogous to the occurence of spatial
dispersion in crystal optics \cite{LandauLifshitz}, the
hole array acting as a 2D crystal. \\
\hspace*{5mm}In this situation, a space-variant output SOP is
readily expected. Since the SPs are dominantly longitudinal
surface waves, the excited areas on the hole array extending
beyond the excitation spot are mainly linearly polarized along the
axes of the reciprocal lattice assigned to the excited SP mode.
Polarization isotropy is found only in the excitation spot itself
whereas a progressive modification of the input polarization is
involved in the propagation of SPs, towards linear polarizations
oriented along axes of the reciprocal lattice. In the
Fourier-related far field, as sketched in Fig. (1), the output SOP
contains therefore various angularly separated polarization
components, i.e. it is (partly) depolarized. Most essential for
our analysis is that, due to the far field angle dependence of the
transmission matrix amplitudes $t_{ij}$, symmetry operations of
the chosen array must be carried out within the angular
integration, bringing forth {\it different} constraints on ${\cal
P}$ for the case of a rotationally symmetric beam versus a plane
wave as input illumination.\\
\hspace*{5mm}More specifically, spatial symmetry operations are
easily analyzed by referring to the circular-state basis
$(\sigma_{+},\sigma_{-})$. Mirror symmetry exchanges $\sigma_{+}$
into $\sigma_{-}$ and vice-versa and is directly related to
complex conjugation. Rotation simply corresponds to phase factors
induced on transformed elements. As a $4^{\rm th}$-rank tensor,
${\cal P}$ features in this basis {\it a priori} $16$ independent
components. Mirror symmetry in combination with angular
integration reduces this to $8$ independent ones and moreover
implies that these $8$ components are real-valued. Then, four- and
six-fold rotational invariance retain only $4$ independent
components for the former and $3$ for the latter. Restricting the
${\cal P}$ tensor to these $4$ and $3$ components, rotational
symmetries together with mirror symmetry force the ${\bf
\sigma}_{\alpha}$ and ${\bf \sigma}_{\beta} $ components to the
tight $\delta_{\alpha\beta}$ ``selection-rule''. This entails that
Mueller matrices for square and hexagonal arrays must be {\it
diagonal}, with the $4$ on-diagonal elements given in the $(x,y)$
laboratory frame as
\begin{eqnarray}
{\rm M}_{0} &=& \left\langle|t_{xx}|^{2}+|t_{yy}|^{2}
+|t_{xy}|^{2}+|t_{yx}|^{2}\right\rangle  \nonumber  \\
{\rm M}_{1}  &=& \left\langle|t_{xx}|^{2}+|t_{yy}|^{2}
-|t_{xy}|^{2}-|t_{yx}|^{2}\right\rangle   \nonumber  \\
{\rm M}_{2}  &=&  2\Re \left[\left\langle t_{xx}t_{yy}^{\star} +
t_{xy}t_{yx}^{\star} \right\rangle \right] \nonumber  \\
{\rm M}_{3} &=&  2\Re \left[\left\langle t_{xx}t_{yy}^{\star}
-t_{xy}t_{yx}^{\star} \right\rangle \right],
\end{eqnarray}
omitting $\lambda$ and $\theta$ dependences. In practice, the
transmission matrix components $t_{ij}$ depend on the actual
structure of the chosen array, in particular on the dimensions of
the holes and on the permittivity of the metal used (real and
imaginary parts). These parameters characterize in fact the SP
modes on the array interfaces; their quantitative evaluation (i.e.
beyond symmetry aspects) requires extensive model calculations
that are outside the scope of this Letter. As special cases, we
find that the Mueller matrix ${\bf M}$ is proportional to the
identity matrix both in the limit of zero hole size and that of
highly dissipative metal permittivity (for e.g. Cr) where SPs
cannot be excited. \\
\hspace*{5mm}The diagonal forms of the Mueller matrices for square
and hexagonal arrays immediately reveal that there is no mixing of
Stokes parameters so that the excitation does not produce any new
spatially averaged polarization in the transmitted beam. With no
enpolarizing capabilities \cite{LuOptComm1998}, these arrays act
as purely depolarizing optical elements. To each input Stokes
parameter $i=(1,3)$ is associated a given degree of polarization
$\Pi_{i}^{\rm out}={\rm M}_{i} / {\rm M}_{0}<1$, which depends on
the opening angle of the input beam. For an hexagonal array, the
reduction from $4$ to $3$ independent ${\cal P}$ components
results in $3$ independent elements of the Mueller matrix with
${\rm M}_{1}={\rm M}_{2}$. As a consequence, two input beams with
the different pure SOPs ${\bf S}^{\rm in}/{\rm S}_{0}^{\rm in}=
(1,\pm 1,0,0)$ and ${\bf S}^{\rm in} /{\rm S}_{0}^{\rm
in}=(1,0,\pm 1,0)$ will suffer the same amount of depolarization
in transmission. Due to the absence of mixing of Stokes
parameters, this can be measured directly through standard crossed
polarization analysis. It is easy to show that such a direct
measurement of depolarization is no longer possible for a
rectangular array due to the loss of the simple diagonal structure
of the Mueller matrix when going to two-fold
symmetry \cite{GenetInPrep}. \\
\hspace*{5mm}In conclusion, we have reported a symmetry-based
theory of the optical polarization properties of nanofabricated
metal hole arrays in the presence of SPs excited on the arrays
interfaces by a finite-diameter beam. Our theory has two essential
ingredients: (i) multi-plane wave illumination and (ii) spatial
dispersion related to SP excitations. The former corresponds to a
far field angular integration performed at the level of ${\cal P}$
components, and the latter implies that this integration is done
over transmission amplitudes depending actually on far field
angles. A tensorial symmetry analysis shows that such arrays
induce depolarization in transmission, a fundamental property
which has not been addressed before. We stress that our analysis
is based on the assumption of a fully polarized input SOP. If the
input beam itself is already partially depolarized, that is if
$\Pi^{\rm in} <1$, the angular integration cannot be performed
separately on the Mueller matrix ${\bf M}$ and on the input Stokes
vector ${\rm S}^{\rm in}$, bringing additional complexity into the
problem. Additionally, and in view of the recent discovery of
SP-assisted polarization entanglement
\cite{AltewischerNature2002}, it would be very interesting to
develop a quantum version of our approach which could  be based
upon {\it
twin-photon} Stokes parameters \cite{AbouraddyOptComm2002}. \\
\hspace*{5mm}As a further outlook, we have started a theoretical
and experimental extension of the present work to depolarization
effects with non-circular holes. Also quasiperiodic structures,
such as 2D Penrose quasicrystals \cite{SteinhardtPNAS1996}, are of
particular interest; there, the appearance of forbidden point
group symmetries should lead to
unexpected depolarization behaviours. \\

This work has been supported by the Stichting voor Fundamenteel
Onderzoek der Materie (FOM) and by the European Union under the
IST - ATESIT contract.

\end{document}